\documentclass[12pt]{article}

\usepackage{graphicx}
\usepackage{float}

\def\gtwid{\mathrel{\raise.3ex\hbox{$>$\kern-.75em\lower1ex\hbox{$\sim$}}}}
\def\ltwid{\mathrel{\raise.3ex\hbox{$<$\kern-.75em\lower1ex\hbox{$\sim$}}}}
\def\square{\kern1pt\vbox{\hrule height 1.2pt\hbox{\vrule width 1.2pt\hskip 3pt
   \vbox{\vskip 6pt}\hskip 3pt\vrule width 0.6pt}\hrule height 0.6pt}\kern1pt}

\begin{document}

\begin{titlepage}

\begin{flushright}
UFIFT-QG-19-01
\end{flushright}

\vskip 0.2cm

\begin{center}
{\bf Ricci Subtraction for Cosmological Coleman-Weinberg Potentials}
\end{center}

\vskip 0.2cm

\begin{center}
S. P. Miao$^{1*}$, S. Park$^{2\star}$ and R. P. Woodard$^{3\dagger}$
\end{center}

\begin{center}
\it{$^{1}$ Department of Physics, National Cheng Kung University, \\
No. 1 University Road, Tainan City 70101, TAIWAN}
\end{center}

\begin{center}
\it{$^{2}$ CEICO, Institute of Physics of the Czech Academy of Sciences, \\ 
Na Slovance 2, 18221 Prague 8 CZECH REPUBLIC}
\end{center}

\begin{center}
\it{$^{3}$ Department of Physics, University of Florida,\\
Gainesville, FL 32611, UNITED STATES}
\end{center}

\vspace{0.2cm}

\begin{center}
ABSTRACT
\end{center}
We reconsider the fine-tuning problem of scalar-driven inflation arising from
the need to couple the inflaton to ordinary matter in order to make reheating
efficient. Quantum fluctuations of this matter induce Coleman-Weinberg 
corrections to the inflaton potential, depending (for de Sitter background) in 
a complex way on the ratio of the inflaton to the Hubble parameter. These corrections
are not Planck-suppressed and cannot be completely subtracted because they are not
even local for a general geometry. A previous study showed that it is not satisfactory 
to subtract a local function of just the inflaton and the {\it initial} Hubble 
parameter. This paper examines the other allowed possibility of subtracting a local 
function of the inflaton and the Ricci scalar. The problem in this case is that the 
new, scalar degree of freedom induced by the subtraction causes inflation to end
almost instantly.

\begin{flushleft}
\end{flushleft}

\vspace{0.2cm}

\begin{flushleft}
$^{*}$ e-mail: spmiao5@mail.ncku.edu.tw \\
$^{\star}$ email:  park@fzu.cz \\
$^{\dagger}$ e-mail: woodard@phys.ufl.edu
\end{flushleft}

\end{titlepage}

\section{Introduction}

Primordial inflation driven by the potential energy of a scalar inflaton,
\begin{equation}
\mathcal{L} = \frac{R \sqrt{-g}}{16 \pi G} - \frac12 \partial_{\mu}
\varphi \partial_{\nu} \varphi g^{\mu\nu} \sqrt{-g} - V(\varphi) 
\sqrt{-g} \; , \label{Lagrangian}
\end{equation}
suffers from many fine tuning problems \cite{Akrami:2018odb}. These include 
the need to make the potential very flat, the need to choose very special 
initial conditions to make inflation start, and the need to keep inflation 
predictive by avoiding the formation of a multiverse \cite{Ijjas:2013vea}.
The implications of the increasingly stringent upper limits on the 
tensor-to-scalar ratio have caused some of the pioneers of inflation to 
question its testability \cite{Guth:2013sya,Linde:2014nna,Ijjas:2014nta}.

This paper is aimed at a different sort of fine tuning problem which may 
prove equally serious: the Coleman-Weinberg corrections \cite{Coleman:1973jx} 
to the inflaton potential that are generated when ordinary matter is coupled 
to the inflaton to facilitate re-heating. These corrections are too large to
be ignored because they are not Planck-suppressed \cite{Green:2007gs}. The
usual assumption has been that Coleman-Weinberg corrections are local functions
of the inflaton which could be completely subtracted off, however, it has 
recently been shown that cosmological Coleman-Weinberg corrections depend 
{\it nonlocally} on the metric \cite{Miao:2015oba}, which precludes their
complete subtraction. 

There are two possible local subtraction schemes:
\begin{enumerate}
\item{Subtract a local function of the inflaton which exactly cancels the
cosmological Coleman-Weinberg potential at the beginning of inflation; or}
\item{Subtract a local function of the inflaton and the Ricci scalar which
exactly cancels the cosmological Coleman-Weinberg potential when the first
slow roll parameter vanishes.}
\end{enumerate}
A recent study of the first possibility concluded that it is not viable
\cite{Liao:2018sci}. When the inflaton is coupled to fermions, inflation never
ends unless the coupling constant is chosen so small as to endanger re-heating,
and then an initial reduction of the expansion rate still results in de Sitter
expansion at a lower rate. When a charged inflaton is coupled to gauge bosons,
inflation ends almost immediately, again unless the coupling constant is chosen
so small as to endanger re-heating.

The purpose of this paper is to study the second possible subtraction scheme.
Section 2 details the form of cosmological Coleman-Weinberg potentials for
fermionic and for bosonic couplings. Section 3 gives the evolution equations
associated with the second subtraction scheme. The effect on the simple 
$V(\varphi) = \frac12 m^2 \varphi^2$ model is worked out for fermion and boson
couplings in section 4. We discuss the results in section 5

\section{Review of Past Work on the Problem}

The purpose of this section is to explain the cosmological Coleman-Weinberg
corrections from fermions and from gauge bosons, which differ profoundly from the
simple $\mp \varphi^4 \ln(\varphi)$ form that pertains in flat space \cite{Coleman:1973jx}. 
The section begins by reviewing explicit results from computations in de Sitter 
background. We then explain our assumption for how to generalize these de Sitter
results to a general spatially flat, homogeneous and isotropic background geometry,
\begin{equation}
ds^2 = -dt^2 + a^2(t) d\vec{x} \!\cdot\! d\vec{x} \qquad \Longrightarrow \qquad H(t) 
\equiv \frac{\dot{a}}{a} \quad , \quad \epsilon(t) \equiv -\frac{\dot{H}}{H^2} \; .
\label{FRW}
\end{equation}
The section closes with a discussion of the Ricci subtraction scheme.

\subsection{Fermion Corrections on de Sitter}

Explicit results have so far only been obtained for de Sitter background, which 
corresponds to $\epsilon = 0$, with $H$ exactly constant. Suppose the inflaton 
$\varphi$ is Yukawa-coupled to a massless, Dirac fermion on this background via 
the interaction $\mathcal{L}_{\rm Yukawa} = -\lambda \varphi \overline{\psi} \psi 
\sqrt{-g}$. The one loop correction to the inflaton effective potential on de Sitter
was originally derived by Candelas and Raine in 1975 \cite{Candelas:1975du}. Of 
course their result depends slightly on conventions of regularization and 
renormalization. Our more recent computation \cite{Miao:2006pn} employed dimensional 
regularization in $D$ spacetime dimensions with conformal and quartic counterterms,
\begin{equation}
\Delta \mathcal{L}^{f} = -\frac12 \delta \xi^{f} \varphi^2 R \sqrt{-g} - \frac1{4!} 
\delta \lambda^{f} \varphi^4 \sqrt{-g} \; . \label{counterLF}
\end{equation}
To simplify the result we took the dimensional regularization scale $\mu$ to be proportional
to the constant Hubble parameter of de Sitter,
\begin{eqnarray}
\delta \xi^{f}_0 & = & \frac{4 \lambda^2 H^{D-4}}{(4 \pi)^{\frac{D}2}} \Biggl\{ 
\frac{\Gamma(1 \!-\! \frac{D}2)}{D (D \!-\! 1)} + \frac{(1 \!-\! \gamma)}{6} + 
O(D \!-\! 4)\Biggr\} \; , \label{dxi0ferm} \\
\delta \lambda^{f}_0 & = & \frac{24 \lambda^4 H^{D-4}}{(4\pi)^{\frac{D}2}} \Biggl\{ 
\Gamma\Bigl(1 \!-\! \frac{D}2\Bigr) + 2 \zeta(3) - 2 \gamma + O(D \!-\! 4)\Biggr\} \; , 
\label{dlambda0ferm}
\end{eqnarray}
where $\gamma = 0.577...$ is the Euler-Mascheroni constant. These choices result in a 
cosmological Coleman-Weinberg potential of the form $\Delta V^{f}_0(\varphi,H) = -
\frac{H^4}{8 \pi^2} \times f(z)$, where $z \equiv \frac{\lambda \varphi}{H}$, and the 
function $f(z)$ is,
\begin{equation}
f(z) = 2 \gamma z^2 - [\zeta(3) \!-\! \gamma] z^4 + 2 \!\! \int_{0}^{z} \!\!\!\! dx \, 
(x \!+\! x^3) \Bigl[ \psi(1 \!+\! i x) + \psi(1 \!-\! ix)\Bigr] \; , \label{f0Fermion}
\end {equation}
and $\psi(x) \equiv \frac{d}{dx} \ln[\Gamma(x)]$ is the digamma function.

We assume that de Sitter results can be extended to general homogeneous and siotropic 
geometries (\ref{FRW}) by simply replacing the constant de Sitter Hubble parameter with
the time dependent $H(t)$ for a general background. However, we must be careful to keep 
the dimensional regularization scale constant, which amounts to a small change of the
counterterms (\ref{dxi0ferm}-\ref{dlambda0ferm}),
\begin{eqnarray}
\delta \xi^{f}_1 & = & \frac{4 \lambda^2 \mu^{D-4}}{(4 \pi)^{\frac{D}2}} \Biggl\{ 
\frac{\Gamma(1 \!-\! \frac{D}2)}{D (D \!-\! 1)} + \frac{(1 \!-\! \gamma)}{6} + 
O(D \!-\! 4)\Biggr\} \; , \label{dxi1ferm} \\
\delta \lambda^{f}_1 & = & \frac{24 \lambda^4 \mu^{D-4}}{(4\pi)^{\frac{D}2}} \Biggl\{ 
\Gamma\Bigl(1 \!-\! \frac{D}2\Bigr) + 2 \zeta(3) - 2 \gamma + O(D \!-\! 4)\Biggr\} \; . 
\label{dlambda1ferm}
\end{eqnarray}
The net effect is to change $\Delta V^{f}_0(\varphi,H)$ to,
\begin{equation}
\Delta V^{f}_1(\varphi,H) = -\frac{H^4}{8 \pi^2} \Biggl\{ f(z) + z^2 \ln\Bigl(
\frac{H^2}{\mu^2}\Bigr) + \frac12 z^4 \ln\Bigl(\frac{H^2}{\mu^2} \Bigr) \Biggr\} \; . 
\label{V1Fermion}
\end{equation}

\subsection{Gauge Boson Corrections on de Sitter}

The contribution of a gauge boson to a charged inflaton,
\begin{equation}
\mathcal{L}_{\rm vector} = -\Bigl(\partial_{\mu} + i e A_{\mu}\Bigr) \varphi^* 
\Bigl(\partial_{\nu} - i e A_{\nu}\Bigr) \varphi g^{\mu\nu} \sqrt{-g} \; , \label{Lgauge}
\end{equation}
was originally computed on de Sitter background using mode sums by Allen in 1983 
\cite{Allen:1983dg}. As always, the precise result depends on conventions of regularization 
and renormalization. Our more recent, dimensionally regulated computation 
\cite{Prokopec:2007ak} employed the massive photon propagator \cite{Tsamis:2006gj}
with conformal and quartic counterterms,
\begin{equation}
\Delta \mathcal{L}^{b} = -\delta \xi^{b} \varphi^* \varphi R \sqrt{-g} - \frac14 \delta 
\lambda^{b} (\varphi^* \varphi)^2 \sqrt{-g} \; .
\end{equation}
We again chose the dimensional regularization mass scale $\mu$ to be proportional to the
constant de Sitter Hubble parameter,
\begin{eqnarray}
\delta \xi^{b}_0 & = & \frac{e^2 H^{D-4}}{(4 \pi)^{\frac{D}2}} \Biggl\{ \frac1{4 \!-\! D}
+ \frac12 \gamma + O(D \!-\! 4)\Biggr\} \; , \label{dxi0boson} \\
\delta \lambda^{b}_0 & = & \frac{D (D \!-\! 1) e^4 H^{D-4}}{(4\pi)^{\frac{D}2}} \Biggl\{ 
\frac{2}{4 \!-\! D} + \gamma - \frac32 + O(D \!-\! 4)\Biggr\} \; , \label{dlambda0boson}
\end{eqnarray}
These choices result in a cosmological Coleman-Weinberg potential of the form $\Delta 
V^{b}_0(\varphi^* \varphi,H^2) = +\frac{3 H^4}{8 \pi^2} \times b(z)$, where $z \equiv 
\frac{e^2 \varphi^* \varphi}{H^2}$, and $b(z)$ is,
\begin{eqnarray}
\lefteqn{b(z) = \Bigl(-1 \!+\! 2 \gamma \Bigr) z + \Bigl(-\frac32 \!+\! \gamma\Bigr)
z^2 } \nonumber \\
& & \hspace{2cm} + \! \int_{0}^{z} \!\!\!\! dx \, (1 \!+\! x) \Biggl[ 
\psi\Bigl(\frac32 \!+\! \frac12 \sqrt{1 \!-\! 8x} \, \Bigr) + 
\psi\Bigl(\frac32 \!-\! \frac12 \sqrt{1 \!-\! 8x} \, \Bigr)\Biggr] \; . \qquad
\label{f0Boson}
\end {eqnarray}

When generalizing the constant de Sitter Hubble parameter to the time-dependent one of
a general homogeneous and isotropic geometry we must revise the counterterms 
(\ref{dxi0boson}-\ref{dlambda0boson}) to keep the mass scale of dimensional regularization
strictly constant,
\begin{eqnarray}
\delta \xi^{b}_1 & = & \frac{e^2 \mu^{D-4}}{(4 \pi)^{\frac{D}2}} \Biggl\{ \frac1{4 \!-\! D}
+ \frac12 \gamma + O(D \!-\! 4)\Biggr\} \; ,
\label{dxiboson1} \\
\delta \lambda^{b}_1 & = & \frac{D (D \!-\! 1) e^4 \mu^{D-4}}{(4\pi)^{\frac{D}2}} \Biggl\{ 
\frac{2}{4 \!-\! D} + \gamma - \frac32 + O(D \!-\! 4)\Biggr\} \; . \label{dlambda1boson}
\end{eqnarray}
The net effect is to change $\Delta V^{b}_0(\varphi^* \varphi,H^2)$ to,
\begin{equation}
\Delta V^{b}_1(\varphi^* \varphi,H^2) = +\frac{3 H^4}{8 \pi^2} \Biggl\{ b(z) + z 
\ln\Bigl(\frac{H^2}{\mu^2}\Bigr) + \frac12 z^2 \ln\Bigl(\frac{H^2}{\mu^2} \Bigr) \Biggr\} 
\; . \label{V1Boson}
\end{equation}

\subsection{Ricci Subtraction}

Ricci subtraction amounts to subtracting the primitive contribution with the replacement
$H^2(t) \rightarrow \frac1{12} R(t)$. In a homogeneous and isotropic geometry this can
be thought of as an $\epsilon$-dependent Hubble parameter $\overline{H}(t)$,
\begin{equation}
R(t) = 12 H^2(t) + 6 \dot{H}(t) = 12 \Bigl[1 - \frac12 \epsilon(t)\Bigr] H^2(t) \equiv
12 \overline{H}^2(t) \; . \label{Hbardef}
\end{equation} 
Because the quartic terms cancel between the primitive potential and the Ricci subtraction,
the full fermionic result takes the form $\Delta V^{f}_2(\varphi,H) - \Delta 
V^{f}_2(\varphi,\overline{H})$ where,
\begin{eqnarray}
\Delta V^{f}_2(\varphi,H) & \!\!=\!\! & -\frac{H^4}{8\pi^2} \Biggl\{ \Delta f(z) + z^2 
\ln\Bigl(\frac{H^2}{H^2_{\rm inf}}\Bigr) \Biggr\} \quad , \quad z \equiv \frac{\lambda 
\varphi}{H} \; , \label{V2Fermion} \\
\Delta f(z) & \!\!=\!\! & 2 \gamma z^2 - \frac12 z^4 \ln(z^2) + 2 \!\! \int_{0}^{z} \!\!\!\! 
dx \, (x \!+\! x^3) \Bigl[ \psi(1 \!+\! i x) + \psi(1 \!-\! ix)\Bigr] \; . \qquad
\label{DeltafFermion}
\end{eqnarray}
Note that we have chosen the constant mass scale to be the Hubble parameter at the 
beginning of inflation, $\mu = H_{\rm inf}$. The full bosonic result takes the form
$\Delta V^{b}_2(\varphi^* \varphi,H^2) - \Delta V^{b}_2(\varphi^* \varphi,\overline{H}^2)$
where,
\begin{eqnarray}
\Delta V^{b}_2(\varphi^* \varphi,H^2) & = & +\frac{3 H^4}{8\pi^2} \Biggl\{ \Delta b(z) 
+ z \ln\Bigl(\frac{H^2}{H^2_{\rm inf}}\Bigr) \Biggr\} \quad , \quad z \equiv 
\frac{e^2 \varphi^* \varphi}{H^2} \; , \label{V2Boson} \\
\Delta b(z) & = & \Bigl(-1 \!+\! 2 \gamma\Bigr) z - \frac12 z^2 \ln(2 z) \nonumber \\
& & \hspace{-0.5cm} + \!\! \int_{0}^{z} \!\!\!\! dx \, (1 \!+\! x) \Biggl[ \psi\Bigl(\frac32
\!+\! \frac12 \sqrt{1 \!-\! 8 x} \, \Bigr) + \psi\Bigl(\frac32 \!-\! \frac12 \sqrt{1 \!-\!
8 x} \, \Bigr)\Biggr] \; . \qquad\label{DeltabBoson}
\end{eqnarray} 

\section{The Modified Evolution Equations}

The purpose of this section is to work out the two Friedmann equations for the
case in which the cosmological Coleman-Weinberg potential depends on the 
Hubble parameter, and it is subtracted by a function which depends on the
Ricci scalar. We also change to dimensionless dependent and independent 
variables.

It is useful to change the evolution variable from co-moving time $t$
to the number of e-foldings from the {\it beginning} of inflation,
\begin{equation}
n \equiv \ln\Bigl( \frac{a(t)}{a(t_i)}\Bigr) \quad \Longrightarrow \quad 
\frac{d}{dt} = H \frac{d}{d n} \quad , \quad \frac{d^2}{d t^2} = H^2 
\Bigl[ \frac{d^2}{d n^2} - \epsilon \frac{d}{d n}\Bigr] \; .
\end{equation}
It is also useful to make the dependent variables dimensionless,
\begin{equation}
\phi(n) \equiv \sqrt{8 \pi G} \times \varphi(t) \qquad , \qquad \chi(n)
\equiv \sqrt{8 \pi G} \times H(t) \; .
\end{equation}
With these variables the slow roll approximation to the (already 
dimensionless) scalar power spectrum becomes,
\begin{equation}
\Delta^2_{\mathcal{R}} \simeq \frac{G H^2}{\pi \epsilon} = \frac1{8 \pi^2}
\frac{\chi^2}{\epsilon} \; . 
\end{equation}
Finally, it is natural to use a dimensionless potential and mass parameter,
\begin{equation}
U \equiv (8 \pi G)^2 \!\times\! V \qquad , \qquad k^2 \equiv 8\pi G 
\!\times\! m^2 \; . 
\end{equation}

The simplest way of expressing the modified field equations is to
imagine that the dimensionless form of the classical potential plus 
the primitive Coleman-Weinberg potential takes the form $U(\phi,\chi)$. The 
Ricci-subtraction takes the similar form $U_{\rm sub}(\phi,
\overline{\chi})$, where we define,
\begin{equation}
\overline{\chi} \equiv \sqrt{1 - \frac12 \epsilon} \, \chi =
\sqrt{\chi^2 + \frac12 \chi \chi'} \; . \label{chidef}
\end{equation}
The two potentials enter the scalar evolution equation the same way,
\begin{equation}
\chi^2 \Bigl[ \phi'' + (3 \!-\! \epsilon) \phi'\Bigr] + 
\frac{\partial U}{\partial \phi} + \frac{\partial U_{\rm sub}}{\partial \phi}
= 0 \; . \label{scaleeqn}
\end{equation}
However, the fact that the subtracted potential $U_{\rm sub}$ depends
upon $\epsilon$, in addition to $\chi$, makes the form of its contributions
to the gravitational field equations very different. The 1st Friedmann 
equation becomes,
\begin{equation}
3 \chi^2 = \frac12 \chi^2 {\phi'}^2 + U - \chi 
\frac{\partial U}{\partial \chi} 
+ U_{\rm sub} -\frac12 (1 \!-\! \epsilon) \chi^2 
\frac{\partial U_{\rm sub}}{\partial \overline{\chi}^2} + \frac12
\chi^2 \frac{d}{dn} \frac{\partial U_{\rm sub}}{\partial 
\overline{\chi}^2} \; . \label{FE1}
\end{equation}
The 2nd Friedmann equation is,
\begin{eqnarray}
\lefteqn{-(3 \!-\! 2 \epsilon) \chi^2 = \frac12 \chi^2 {\phi'}^2 - U +
\chi \frac{\partial U}{\partial \chi} + \frac13 \chi \frac{d}{d n} 
\frac{\partial U}{\partial \chi} } \nonumber \\
& & \hspace{2cm} - U_{\rm sub} + \frac12 \Bigl(1 \!-\! \frac13 \epsilon\Bigr)
\chi^2 \frac{\partial U_{\rm sub}}{\partial \overline{\chi}^2} - \frac16
\chi^2 \Bigl[ \frac{d}{d n} \!+\! 2 \!-\! \epsilon\Bigr] \frac{d}{d n}
\frac{\partial U_{\rm sub}}{\partial \overline{\chi}^2} \; . \label{FE2} 
\qquad 
\end{eqnarray}

One consequence of the final term in equation (\ref{FE1}) is that the first
Friedmann equation involves second derivatives of $\chi(n)$. To see this,
use the chain rule to exhibit the implicit higher derivatives,
\begin{equation}
\frac12 \chi^2 \frac{d}{d n} \frac{\partial U_{\rm sub}}{\partial \overline{\chi}^2}
= \frac12 \chi^2 \Biggl\{ \phi' \frac{\partial^2 U_{\rm sub}}{\partial \phi 
\partial \overline{\chi}^2} + \Bigl[2 \chi \chi' + \frac12 {\chi'}^2 + \frac12
\chi \chi'' \Bigr] \frac{\partial^2 U_{\rm sub}}{\partial \overline{\chi}^4} \Biggr\} .
\label{chainrule}
\end{equation}
Recalling that $\epsilon = -\chi'/\chi$ allows us to express the first 
Friedmann equation (\ref{FE1}) as,
\begin{eqnarray}
\lefteqn{\epsilon' = -4 \epsilon + 2 \epsilon^2 + \frac{4}{\chi^4 \frac{\partial^2 
U_{\rm sub}}{\partial \overline{\chi}^4}} \Biggl\{ -\chi^2 \Bigl[3 \!-\! \frac12
{\phi'}^2\Bigr] + U + U_{\rm sub}} \nonumber \\
& & \hspace{3.5cm} - \chi^2 \Biggl[ 2 \frac{\partial U}{\partial \chi^2} + \frac12
(1 \!-\! \epsilon) \frac{\partial U_{\rm sub}}{\partial \overline{\chi}^2} \Biggr]
+ \frac12 \chi^2 \phi' \frac{\partial^2 U_{\rm sub}}{\partial \phi \partial 
\overline{\chi}^2} \Biggr\} . \label{epsilonevo} \qquad
\end{eqnarray}

The natural initial conditions derive from the slow roll solutions 
for the purely classical model ($U = \frac12 k^2 \phi^2$ and $U_{\rm sub} = 0)$,
\begin{equation}
\phi^2(n) \simeq \phi^2(0) - 4 n \;\; , \;\; \chi^2(n) \simeq \frac16 k^2 
\Bigl[ \phi^2(0) - 4 n\Bigr] \;\; , \;\; \epsilon(n) \simeq \frac{2}{
\phi^2(0) - 4 n} \; . \label{slowroll}
\end{equation}
Hence we obtain a 2-parameter family of initial conditions based on $\phi(0) 
\equiv \phi_0$ and $k$,
\begin{eqnarray}
\phi(0) = \phi_0 \qquad & , & \qquad \phi'(0) = -\frac{2}{\phi_0} \; , 
\label{IC1} \\
\chi(0) = \frac{k \phi_0}{\sqrt{6 \!-\! (\frac{2}{\phi_0})^2}} \qquad & , & 
\qquad \chi'(0) = -\frac{2 \chi_0}{\phi_0^2} \; . \label{IC2}
\end{eqnarray}
Note that these initial conditions (\ref{IC1}-\ref{IC2}) exactly satisfy the classical 
Friedmann equation $3 \chi_0^2 = \frac12 \chi_0^2 {\phi'_0}^2 + \frac12 k^2 \phi_0^2$ 
and also make the first slow roll parameter $\epsilon_0 = 2/\phi_0^2$ agree with
the slow roll approximation (\ref{slowroll}). 

Using the slow roll approximations (\ref{slowroll}) we see that $\phi_0 = 20$ will
give about 100 total e-foldings of inflation. We can also express the power spectrum 
and the scalar spectral index in terms of the evolving first slow roll parameter 
$\epsilon(n)$,
\begin{eqnarray}
\Delta^2_{\mathcal{R}} & \simeq & \frac1{8\pi^2} \frac{\chi^2}{\epsilon} 
\longrightarrow \frac1{8\pi^2} \frac{k^2}{3 \epsilon^2} \; , \label{DR} \\
1 - n_s & \simeq & 2 \epsilon + \frac{\epsilon'}{\epsilon} \longrightarrow
4 \epsilon \; . \label{ns}
\end{eqnarray}
Of course relations (\ref{DR}) and (\ref{ns}) allow us to determine the
constant $k$ in terms of the measured scalar amplitude $A_s$ and spectral
index $n_s$ \cite{Aghanim:2018eyx},
\begin{equation}
k \simeq \pi (1 \!-\! n_s) \sqrt{\frac32 A_s} \simeq 6.13 \times 10^{-6} \; .
\label{kvalue}
\end{equation}

\section{The Fate of the $m^2 \varphi^2$ Model}

The purpose of this section is to numerically simulate the effect of 
primitive Coleman-Weinberg potentials with Ricci subtraction in the context
of the classical $V_{\rm class} = \frac12 m^2 \varphi^2$ model. We begin
with the case of fermionic corrections, and then discuss bosonic corrections.
The generic problem in each case is that the $\chi''(n)$ terms in the first
Friedmann eqaution (\ref{FE1}) excite a new scalar degree of freedom that
causes inflation to end almost immediately when starting from the classical
initial conditions (\ref{IC1}-\ref{IC2}).

\subsection{Fermionic Corrections}

\begin{figure}[H]
\includegraphics[width=4.6cm,height=4.6cm]{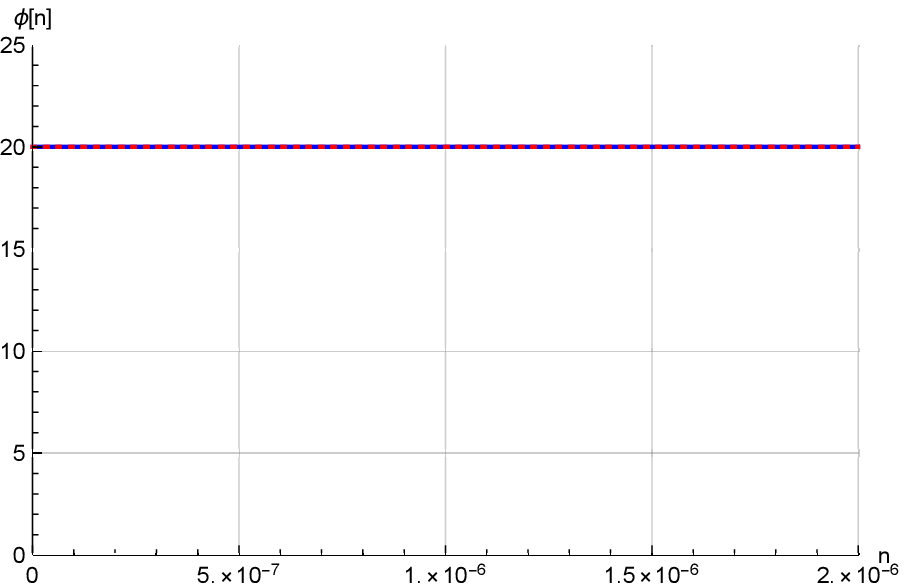}
\hspace{-0.4cm}
\includegraphics[width=4.6cm,height=4.6cm]{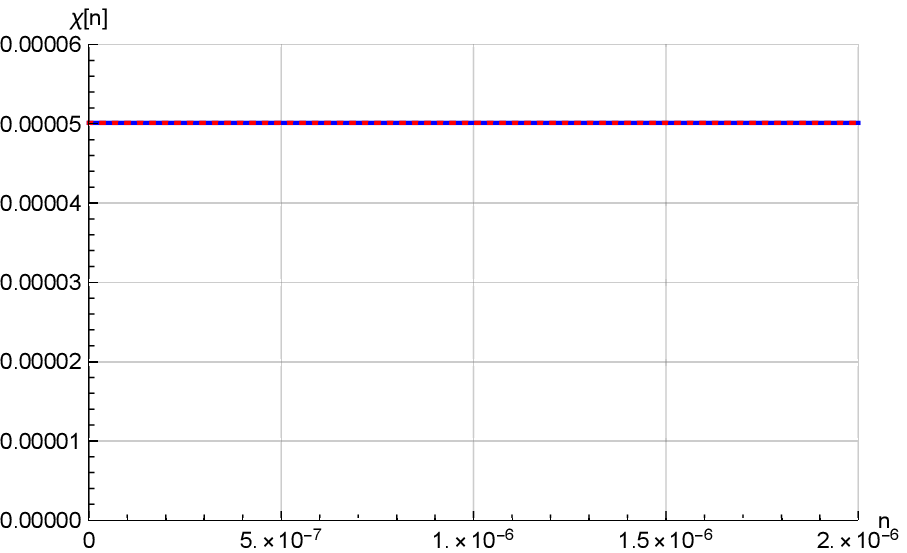}
\hspace{-0.4cm}
\includegraphics[width=4.6cm,height=4.6cm]{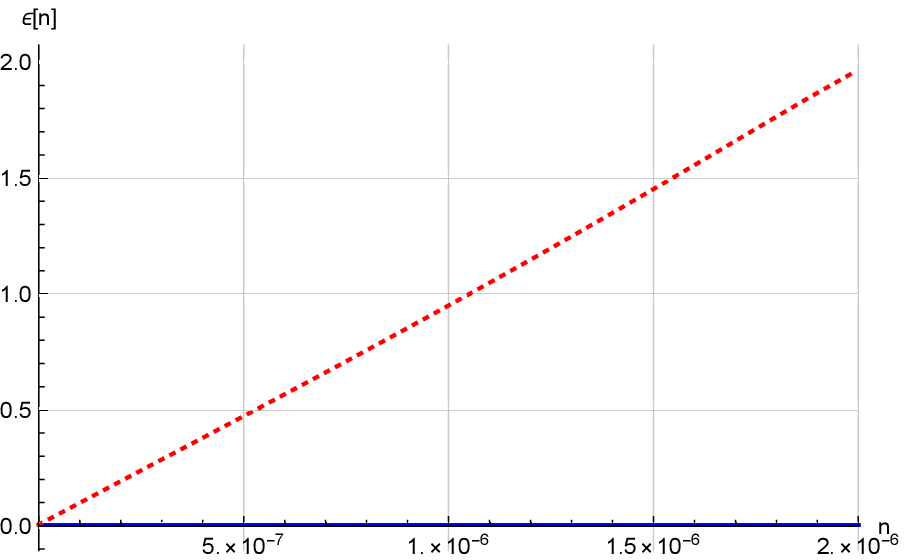}
\caption{Plots of the dimensionless scalar $\phi(n)$ (on the left), the
dimensionless Hubble parameter $\chi(n)$ (middle) and the first slow roll
parameter $\epsilon(n)$ (on the right) for classical model (in blue) and the 
quantum-corrected model (in red dots) with Yukawa coupling $\lambda = 5.5 
\times 10^{-4}$.}
\label{Fermiontry1}
\end{figure}

The Ricci subtraction scheme for fermionic corrections is defined by the
potentials,
\begin{eqnarray}
U^f(\phi,\chi) & = & \frac12 k^2 \phi^2 - \frac{\chi^4}{8 \pi^2} \Biggl[
\Delta f\Bigl( \frac{\lambda \phi}{\chi}\Bigr) + \Bigl( \frac{\lambda \phi}{\chi}
\Bigr)^2 \ln\Bigl( \frac{\chi^2}{\chi^2(0)}\Bigr) \Biggr]\; , \label{Ufermion} \\
U^f_{\rm sub}(\phi,\overline{\chi}) & = & +\frac{\overline{\chi}^4}{8\pi^2}
\Biggl[\Delta f\Bigl( \frac{\lambda \phi}{\overline{\chi}}\Bigr) + \Bigl( 
\frac{\lambda \phi}{\overline{\chi}}\Bigr)^2 \ln\Bigl( 
\frac{\overline{\chi}^2}{\chi^2(0)}\Bigr) \Biggr] \; , \label{Usubfermion}
\end{eqnarray}
where $\Delta f(z)$ was defined in (\ref{DeltafFermion}) and $\overline{\chi} \equiv 
\sqrt{1 - \frac12 \epsilon} \, \chi$.
Figure \ref{Fermiontry1} displays the classical evolution (in blue) versus the 
quantum-corrected model (in red dots) for a moderate coupling of 
$\lambda = 5.5 \times 10^{-4}$.
\noindent While the initial evolution of the scalar and the Hubble parameter is
not visibly affected by the quantum correction, the first slow roll parameter rises
above the inflationary threshold of $\epsilon = 1$ almost immediately.

To understand why Ricci subtraction engenders immediate deviations for $\lambda
= 5.5 \times 10^{-4}$, first note that the initial conditions of the classical model 
(\ref{IC1}-\ref{IC2}) force the initial value of the parameter $z \equiv 
\frac{\lambda \phi}{\chi}$ to be much larger than one,
\begin{equation}
z_0 = \frac{\lambda \phi_0}{\chi_0} = \frac{\lambda}{k} \sqrt{6 - (\frac{2}{\phi_0})^2}
\simeq 220 \; . \label{zzero}
\end{equation}
This means it is valid to use the large $z$ expansion of (\ref{DeltafFermion}) 
\cite{Miao:2015oba},
\begin{equation}
\Delta f(z) = -\frac14 z^4 + z^2 \ln(z^2) - \Bigl( \frac56 \!-\! 2 \gamma\Bigr) z^2
+ \frac{11}{60} \ln(z^2) + O(1) \; . \label{largezf}
\end{equation}
Substituting (\ref{largezf}) in expressions (\ref{Ufermion}-\ref{Usubfermion}) implies,
\begin{eqnarray}
U^{f}(\phi,\chi) & = & \frac12 k^2 \phi^2 - \frac{\chi^4}{8\pi^2} \Biggl\{ -\frac14
\Bigl( \frac{\lambda \phi}{\chi}\Bigr)^4 + \Bigl[ \ln\Bigl( \frac{\lambda^2 \phi^2}{
\chi_0^2}\Bigr) \!-\! \frac56 \!+\! 2 \gamma\Bigr] \Bigl( \frac{\lambda \phi}{\chi}
\Bigr)^2 \nonumber \\
& & \hspace{6cm} + \frac{11}{60} \ln\Bigl( \frac{\lambda^2 \phi^2}{\chi^2}\Bigr) + \dots 
\Biggr\} , \qquad \label{Uflargez} \\
U^f_{\rm sub}(\phi,\overline{\chi}) & = & \frac{\overline{\chi}^4}{8\pi^2} \Biggl\{ -\frac14
\Bigl( \frac{\lambda \phi}{\overline{\chi}}\Bigr)^4 + \Bigl[ \ln\Bigl( \frac{\lambda^2 
\phi^2}{\chi_0^2}\Bigr) \!-\! \frac56 \!+\! 2 \gamma\Bigr] \Bigl( \frac{\lambda 
\phi}{\overline{\chi}} \Bigr)^2 \nonumber \\
& & \hspace{6cm} + \frac{11}{60} \ln\Bigl( \frac{\lambda^2 \phi^2}{\overline{\chi}^2}\Bigr) 
+ \dots \Biggr\} . \qquad \label{Usubflargez} 
\end{eqnarray}
The $(\lambda \phi)^4$ terms cancel out between (\ref{Uflargez}) and (\ref{Usubflargez})
so that the leading contribution usually comes from the $(\lambda \phi)^2$ term,
\begin{eqnarray}
\lefteqn{U^f + U^f_{\rm sub} - \chi^2 \Biggl[ 2 \frac{\partial U^f}{\partial \chi^2} + 
\frac12 (1 \!-\! \epsilon) \frac{\partial U^f_{\rm sub}}{\partial \overline{\chi}^2} 
\Biggr] } \nonumber \\
& & \hspace{3cm} = \frac12 k^2 \phi^2 + \frac{\chi^4}{8 \pi^2} \!\times\! \frac32 z^2 
\Biggl[ \ln\Bigl( \frac{\lambda^2 \phi^2}{\chi_0^2}\Bigr) \!-\! \frac56 \!+\! 2 \gamma\Biggr] 
+ \dots \label{num2} \qquad \\
\lefteqn{\frac12 \chi^2 \phi' \, \frac{\partial^2 U^f_{\rm sub}}{\partial \phi \partial 
\overline{\chi}^2} = \frac{\chi^4}{8 \pi^2} \!\times\! \frac{\phi'}{\phi} z^2 
\Biggl[ \ln\Bigl( \frac{\lambda^2 \phi^2}{\chi_0^2}\Bigr) \!+\! \frac16 \!+\! 2 \gamma\Biggr] 
+ \dots } \label{num3}
\end{eqnarray}
However, the $(\lambda \phi)^2$ term makes no contribution to the denominator, so one must 
go one order higher,
\begin{equation}
\chi^4 \frac{\partial^2 U^f_{\rm sub}}{\partial \overline{\chi}^4} = \frac{\chi^4}{8 \pi^2}
\!\times\! \frac{11}{60} \Biggl[ 2 \ln\Bigl( \frac{\lambda^2 \phi^2}{\overline{\chi}^2} 
\Bigr) - 3 \Biggr] + \dots \label{den}
\end{equation}
Taking account of the fact that the classical terms initially cancel, and that $\phi_0'/\phi_0 
= -\epsilon_0$, the large $z_0$ form of (\ref{epsilonevo}) is,
\begin{equation}
\epsilon_0' \simeq -4\epsilon_0 + 2 \epsilon_0^2 + \frac{6 z_0^2 [\ln(z_0^2) \!-\! \frac56
\!+\! 2 \gamma] - 4 \epsilon_0 z_0^2 [\ln(z_0^2) \!+\! \frac16 \!+\! 2 \gamma]}{\frac{11}{60}
[2 \ln(z_0^2) \!-\! 2 \ln(1 \!-\! \frac12 \epsilon_0) \!-\! 3]} \simeq 9.4 \times 10^5 \; . 
\label{epsilonprimezero}
\end{equation}
This compares with the slow roll result of $\epsilon'_0 = 2 \epsilon_0^2 = 5 \times 10^{-5}$,
and explains why the subtraction term brings inflation to such an abrupt end.

\begin{figure}[H]
\includegraphics[width=10.0cm,height=4.0cm]{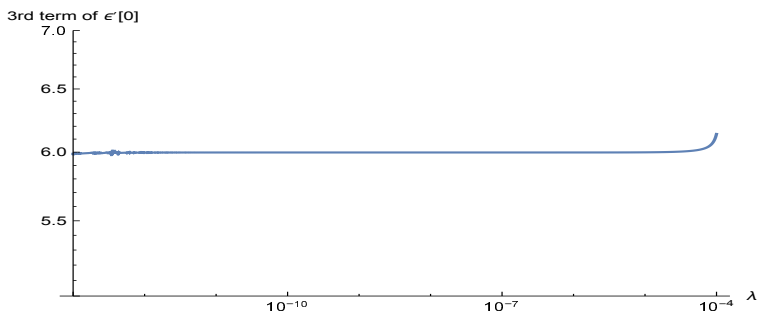}
\caption{Log-log plot of the final term in relation (\ref{epsilonevo}) for
$\epsilon'(0)$ as a function of $\lambda$ in the range $10^{-13} < \lambda
< 10^{-4}$.}
\label{Fermiontry2}
\end{figure}

Because the large fraction in (\ref{epsilonprimezero}) scales like $\lambda^2$ one might 
expect that decreasing $\lambda$ reduces $\epsilon'(0)$. This is indeed true for as long as 
the large $z$ regime pertains, but $\epsilon'(0)$ approaches a constant value of about 6 in 
the small $z$ regime, as can be seen from Figure~\ref{Fermiontry2}.
\noindent The asymptotic limit of $\epsilon'(0) \simeq 6$ is still much too large, 
corresponding to only about one e-folding of inflation.

To understand the small $z$ limit of $\epsilon'(0)$, note first that the small $z$ 
expansion of $\Delta f(z)$ is \cite{Miao:2015oba},
\begin{equation}
\Delta f(z) = -\frac12 z^4 \ln(z^2) + \Bigl[ \zeta(3) - \gamma\Bigr] z^4 + \frac23
\Bigl[ \zeta(3) - \zeta(5)\Bigr] z^6 + O(z^8) \; . 
\end{equation}
Comparison with expressions (\ref{Ufermion}) and (\ref{Usubfermion}) implies that the
small $\lambda$ limiting forms derive from the conformal renormalization,
\begin{eqnarray}
U^f(\phi,\chi) - \frac12 k^2 \phi^2 & \longrightarrow & -\frac{\lambda^2 \phi^2 
\chi^2}{8 \pi^2} \ln\Bigl( \frac{\chi^2}{\chi_0^2}\Bigr) \; , \label{smallUf} \\
U^f_{\rm sub}(\phi,\overline{\chi}) & \longrightarrow & + \frac{\lambda^2 \phi^2 
\overline{\chi}^2}{8 \pi^2} \ln\Bigl( \frac{\overline{\chi}^2}{\chi_0^2}\Bigr) \; .
\label{smallUfsub}
\end{eqnarray}
Substituting (\ref{smallUf}-\ref{smallUfsub}) in the final term of (\ref{epsilonevo})
gives,
\begin{equation}
\epsilon'_0 \longrightarrow -4\epsilon_0 + 2 \epsilon_0^2 + \frac{6 \!-\! 2 \epsilon_0
\!+\! 2 (1 \!-\! 2 \epsilon_0) \ln(1 \!-\! \frac12 \epsilon_0)}{(1 \!-\! \frac12 
\epsilon_0)^{-1}} \; . \label{smallepsprime}
\end{equation}

A striking feature of Figure~\ref{Fermiontry2} and expression (\ref{smallepsprime}) 
is that the limit $\lambda \rightarrow 0$ fails to agree with the case of $\lambda = 0$
for which there is no change to classical inflation. This seems contradictory but is in
fact the standard signature of a perturbation that changes the number of derivatives.
A simple example is the higher derivative extension of the simple harmonic oscillator
considered in section 2.2 of \cite{Woodard:2015zca}. The oscillator's position is $x(t)$ 
and its Lagrangian is,
\begin{equation}
L = -\frac{\epsilon m}{2 \omega^2} \ddot{x}^2 + \frac{m}{2} \dot{x}^2 - 
\frac{m \omega^2}{2} x^2 \; . \label{HDoscillator}
\end{equation}
When $\epsilon = 0$ this system reduces to the simple harmonic oscillator which has 
two pieces of initial value data and whose energy is bounded below. However, for any
nonzero value of $\epsilon$ the system has {\it four} pieces of initial value data,
and its energy is unbounded below. Because the effect of the higher derivative 
perturbation (in our inflation model) never becomes small, no matter how small the 
coupling constant, it follows that perturbation theory breaks down.

\subsection{Corrections from Gauge Bosons}

Making the inflaton complex causes a few small changes in the key equations of section 3.
Because the two potentials $U^b(\phi^* \phi,\chi^2)$ and $U^b_{\rm sub}(\phi^* \phi,
\overline{\chi}^2)$ depend on the norm-squared of the scalar, the evolution equation for
the inflaton becomes,
\begin{equation}
\chi^2 \Bigl[ \phi'' + (3 \!-\! \epsilon) \phi'\Bigr] + \phi \Biggl[ \frac{\partial 
U^b(\phi^* \phi,\chi^2)}{\partial \phi^* \phi} + \frac{\partial U^b_{\rm sub}(\phi^* \phi,
\overline{\chi}^2)}{\partial \phi^* \phi} \Biggr] = 0 \; . \label{bosonscalar}
\end{equation}
The first Friedmann equation takes the form,
\begin{equation}
3 \chi^2 = \chi^2 {\phi'}^* \phi' + U^b + U^b_{\rm sub} - \chi^2 \Biggl[ 2
\frac{\partial U^b}{\partial \chi^2} + \frac12 (1 \!-\! \epsilon) 
\frac{\partial U^b_{\rm sub}}{\partial \overline{\chi}^2} \Biggr] + \frac12 \chi^2
\frac{d}{d n} \frac{\partial U^b_{\rm sub}}{\partial \overline{\chi}^2} \; .
\label{bosonFE1}
\end{equation}
This gives an evolution equation for the first slow roll parameter analogous to
(\ref{epsilonevo}),
\begin{eqnarray}
\lefteqn{\epsilon' = -4 \epsilon + 2 \epsilon^2 + \frac{4}{\chi^4 \frac{\partial^2 
U^b_{\rm sub}}{\partial \overline{\chi}^4}} \Biggl\{ -\chi^2 \Bigl[3 \!-\! {\phi'}^*
\phi'\Bigr] + U^b + U^b_{\rm sub}} \nonumber \\
& & \hspace{2.5cm} - \chi^2 \Biggl[ 2 \frac{\partial U^b}{\partial \chi^2} + \frac12
(1 \!-\! \epsilon) \frac{\partial U^b_{\rm sub}}{\partial \overline{\chi}^2} \Biggr]
+ \frac12 \chi^2 (\phi^* \phi)' \frac{\partial^2 U^b_{\rm sub}}{\partial \phi^* \phi 
\partial \overline{\chi}^2} \Biggr\} . \label{bosonepsilonevo} \qquad
\end{eqnarray}
Even though we do not use it, the second Friedmann equation is,
\begin{eqnarray}
\lefteqn{ -(3 \!-\! 2 \epsilon) \chi^2 = \chi^2 {\phi'}^* \phi' - U^b - U^b_{\rm sub} 
+ \chi^2 \Biggl[ 2 \frac{\partial U^b}{\partial \chi^2} + \frac12 \Bigl(1 \!-\! \frac13 
\epsilon\Bigr) \frac{\partial U^b_{\rm sub}}{\partial \overline{\chi}^2} \Biggr] }
\nonumber \\
& & \hspace{3.5cm} + \frac13 \chi^2 \Bigl[\frac{d}{d n} \!-\! \epsilon\Bigr] 
\frac{\partial U^b}{\partial \chi^2} - \frac16 \chi^2 \Bigl[ \frac{d}{d n} \!+\! 2 \!-\! 
\epsilon\Bigr] \frac{d}{d n} \frac{\partial U^b_{\rm sub}}{\partial \overline{\chi}^2} \; . 
\qquad \label{bosonFE2}
\end{eqnarray}
And the initial values (assuming $\phi_0$ is real) become,
\begin{eqnarray}
\phi(0) = \phi_0 \qquad & , & \qquad \phi'(0) = -\frac{2}{\phi_0} \; , 
\label{bosonIC1} \\
\chi(0) = \frac{k \phi_0}{\sqrt{3 \!-\! (\frac{2}{\phi_0})^2}} \qquad & , & 
\qquad \chi'(0) = -\frac{2 \chi_0}{\phi_0^2} \; . \label{bosonIC2}
\end{eqnarray}
We continue to use $\phi_0 = 20$, with the value of $k$ given in (\ref{kvalue}).

\begin{figure}[H]
\includegraphics[width=4.6cm,height=4.6cm]{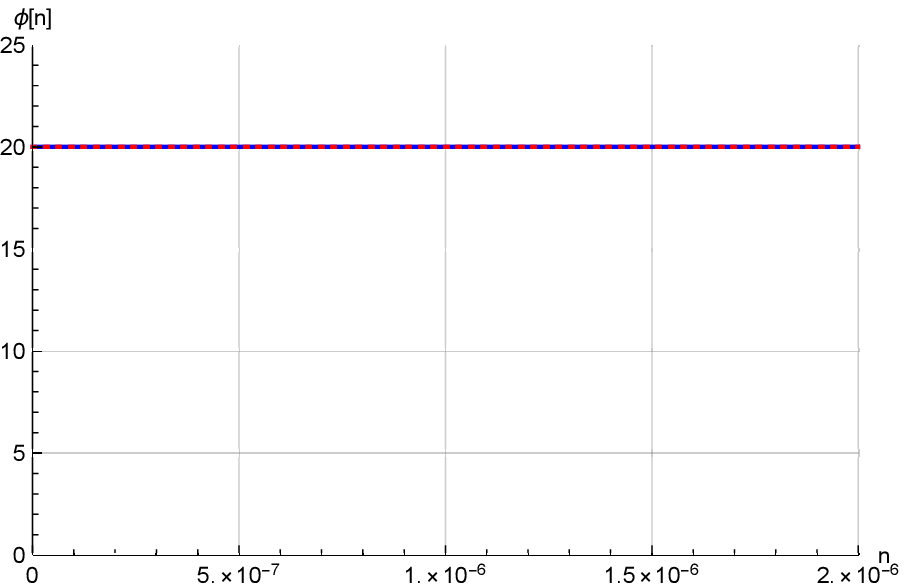}
\hspace{-0.4cm}
\includegraphics[width=4.6cm,height=4.6cm]{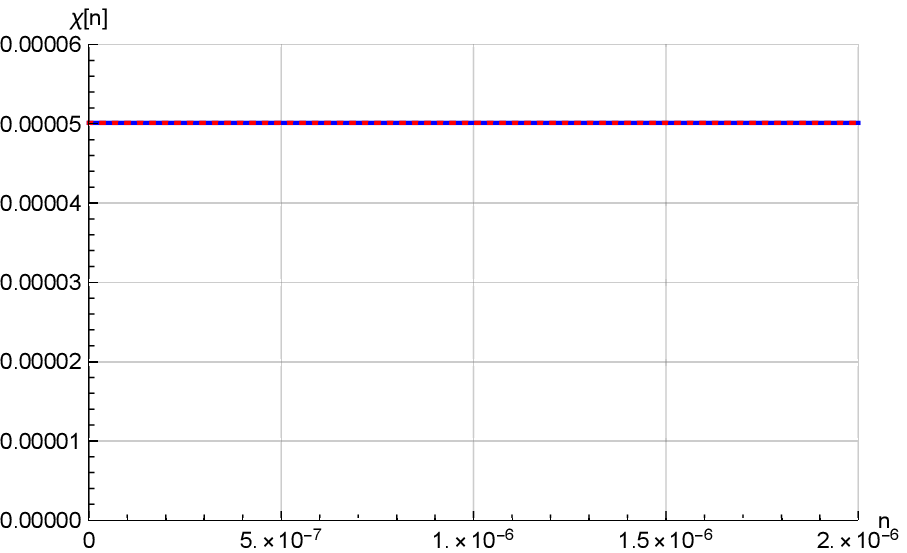}
\hspace{-0.4cm}
\includegraphics[width=4.6cm,height=4.6cm]{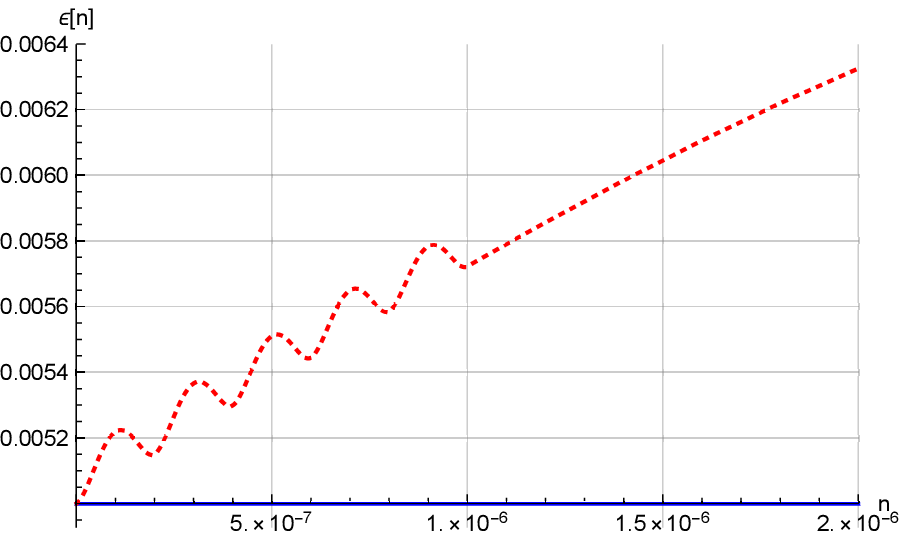}
\caption{Plots of the dimensionless scalar $\phi(n)$ (on the left), the
dimensionless Hubble parameter $\chi(n)$ (middle) and the first slow roll
parameter $\epsilon(n)$ (on the right) for classical model (in blue) and
the quantum-corrected model (in red) with the charge-squared $e^2 = \frac{4\pi}{137} 
\times 10^{-8.5} \simeq 2.9 \times 10^{-10}$.}
\label{Bosontry1}
\end{figure}

The Ricci subtraction scheme for bosons is defined by these potentials, 
\begin{eqnarray}
U^b(\phi^* \phi,\chi^2) & = & k^2 \phi^* \phi + \frac{3 \chi^4}{8 \pi^2} \Biggl[
\Delta b\Bigl( \frac{e^2 \phi^* \phi}{\chi^2}\Bigr) + \frac{e^2 \phi^* \phi}{\chi^2}
\ln\Bigl(\frac{\chi^2}{\chi^2(0)}\Bigr) \Biggr] \; , \label{Uboson} \\
U^b_{\rm sub}(\phi^* \phi,\overline{\chi}^2) & = & -\frac{3 \overline{\chi}^4}{8\pi^2}
\Biggl[ \Delta b\Bigl( \frac{e^2 \phi^* \phi}{\overline{\chi}^2}\Bigr) + 
\frac{e^2 \phi^* \phi}{\overline{\chi}^2} \ln\Bigl(\frac{\overline{\chi}^2}{
\chi^2(0)}\Bigr) \Biggr] \; , \label{Usubboson}
\end{eqnarray}
where $\Delta b(z)$ was defined in (\ref{DeltabBoson}) and $\overline{\chi} \equiv 
\sqrt{1 - \frac12 \epsilon} \, \chi$. Figure \ref{Bosontry1} compares the classical
evolution (in blue) with the quantum-corrected (in red dots) for a charge $e^2 \simeq 
2.9 \times 10^{-10}$ which is three hundred million times weaker than electromagnetism.

The rapid onset of deviations from classical evolution evident in Figure \ref{Bosontry1} 
has the same explanation for bosons as for fermions. Even for the small coupling $e^2 
\simeq 2.9 \times 10^{-10}$ the initial value of $z = \frac{e^2 \phi^* \phi}{\chi^2}$ is
larger than one,
\begin{equation}
z_0 \equiv \frac{e^2 \phi_0^2}{\chi_0^2} = \frac{e^2}{k^2} \Bigl[3 - \frac4{\phi_0^2}
\Bigr] \simeq 23.1 \; . \label{initialzboson}
\end{equation}
Just as for fermions, this means we can simplify relation (\ref{bosonepsilonevo})
using the large $z$ expansion of $\Delta b(z)$ \cite{Miao:2015oba},
\begin{equation}
\Delta b(z) = -\frac14 z^2 + z \ln(2z) - \Bigl( \frac53 \!-\! 2 \gamma\Bigr) z +
\frac{19}{60} \ln(z) + O(1) \; . \label{largezb}
\end{equation}
The corresponding large argument expansions of the potentials are,
\begin{eqnarray}
U^{b}(\phi^* \phi,\chi^2) & = & k^2 \phi^* \phi + \frac{3 \chi^4}{8\pi^2} \Biggl\{ 
-\frac14 \Bigl( \frac{e^2 \phi^* \phi}{\chi^2}\Bigr)^2 \nonumber \\
& & \hspace{-0.3cm} + \Bigl[ \ln\Bigl( \frac{2 e^2 \phi^* \phi}{\chi_0^2}\Bigr) \!-\! 
\frac53 \!+\! 2 \gamma\Bigr] \frac{e^2 \phi^* \phi}{\chi^2} + \frac{19}{60} \ln\Bigl( 
\frac{e^2 \phi^* \phi}{\chi^2}\Bigr) + \dots \Biggr\} , \qquad \label{Ublargez} \\
U^b_{\rm sub}(\phi^* \phi,\overline{\chi}^2) & = & -\frac{3 \overline{\chi}^4}{8\pi^2} 
\Biggl\{ -\frac14 \Bigl( \frac{e^2 \phi^* \phi}{\overline{\chi}^2}\Bigr)^2 \nonumber \\
& & \hspace{-0.3cm} + \Bigl[ \ln\Bigl( \frac{2 e^2 \phi^* \phi}{\chi_0^2}\Bigr) \!-\!
\frac53 \!+\! 2 \gamma\Bigr] \frac{e^2 \phi^* \phi}{\overline{\chi}^2} + \frac{19}{60} 
\ln\Bigl( \frac{e^2 \phi^* \phi}{\overline{\chi}^2}\Bigr) + \dots \Biggr\} . \qquad 
\label{Usubblargez} 
\end{eqnarray}
Just as for fermions, the order $e^2 \phi^* \phi$ terms in (\ref{Ublargez}-\ref{Usubblargez})
make the dominant contributions to the numerator of expression (\ref{bosonepsilonevo}), but
the denominator is a crucial order weaker, 
\begin{equation}
\epsilon_0' \simeq -4\epsilon_0 + 2 \epsilon_0^2 + \frac{6 z_0 [\ln(2 z_0) \!-\! \frac53
\!+\! 2 \gamma] - 4 \epsilon_0 z_0 [\ln(2 z_0) \!-\! \frac23 \!+\! 2 \gamma]}{\frac{19}{60}
[2 \ln(z_0) \!-\! 2 \ln(1 \!-\! \frac12 \epsilon_0) \!-\! 3]} \simeq 441 \; . 
\label{bosonepsilonprimezero}
\end{equation}

\begin{figure}[H]
\includegraphics[width=10.0cm,height=4.0cm]{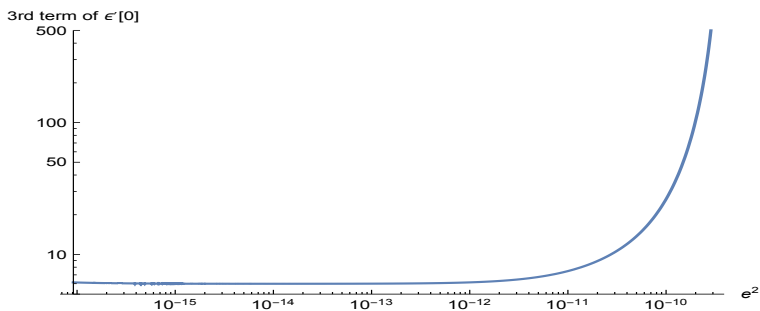}
\caption{Plot of the 3rd term in expression (\ref{bosonepsilonevo}) for $\epsilon'(0)$
as a function of $e^2$.}
\label{Bosontry2}
\end{figure}

Just as we found for fermions, $\epsilon'(0)$ can be decreased by reducing the 
coupling constant, but it eventually approaches a value that is still much too large. 
This can be seen from Figure~\ref{Bosontry2}.
\noindent The analytic derivation follows from the small $e^2$ limiting forms of the 
quantum part of $U^b$ and $U^b_{\rm sub}$,
\begin{eqnarray}
U^b(\phi^* \phi,\chi^2) - k^2 \phi^* \phi & \longrightarrow & 
\frac{3 e^2 \phi^* \phi \chi^2}{8\pi^2} \ln\Bigl( \frac{\chi^2}{\chi_0^2}\Bigr) 
\; , \\
U^b_{\rm sub}(\phi^* \phi,\overline{\chi}^2) & \longrightarrow & 
-\frac{3 e^2 \phi^* \phi \overline{\chi}^2}{8 \pi^2} \ln\Bigl( 
\frac{\overline{\chi}^2}{\chi_0^2} \Bigr) \; .
\end{eqnarray}
The analysis, and even the result, is the same as for fermions. Note that the limit
$e^2 \rightarrow 0$, for which there is always an instability, again fails to agree 
with $e^2 = 0$ model, which is classical inflation.

\section{Discussion}

Cosmological Coleman-Weinberg potentials are induced when the inflaton is 
coupled to ordinary matter, typically to facilitate re-heating. Without 
subtraction, these potentials are disastrous to inflation because they are
far too steep and not Planck-suppressed. If they depended only on the
inflaton it would be straightforward to subtract them but they also involve 
the metric in a deep and profound way. Explicit computations on de Sitter 
background, for fermions \cite{Candelas:1975du,Miao:2006pn} and for vector 
bosons \cite{Allen:1983dg,Prokopec:2007ak}, reveal complicated functions of 
the dimensionless ratio of the coupling constant times the inflaton, all
divided by the Hubble parameter. Indirect arguments show that the constant 
Hubble parameter of de Sitter in this ratio cannot be constant for a general
geometry, nor can it even be local \cite{Miao:2015oba}. That poses a major
obstacle to subtracting away cosmological Coleman-Weinberg potentials 
because only local functions of the inflaton and the Ricci scalar can be
employed \cite{Woodard:2006nt}, and neither can completely subtract the
potentials.

A previous study explored the possibility of subtracting a function of just
the inflaton, chosen to completely cancel the cosmological Coleman-Weinberg
potential at the onset of inflation \cite{Liao:2018sci}. What we found for 
moderate coupling constants is that inflation never ends for the corrections
due to fermions, and it ends too soon for the corrections due to vector bosons.
Making the Yukawa coupling very small results in a nearly classical evolution
until late times, at which point the universe approaches de Sitter with a 
much smaller Hubble parameter. An acceptable evolution can only be obtained
by making the vector boson coupling very small, and this degrades the 
efficiency of re-heating.

This paper studied the other possibility: subtracting a function of the inflaton
and the Ricci scalar. One might think (as we originally hoped) that corrections
for this type of subtraction would be suppressed by the smallness of the first 
slow roll parameter. However, the higher time derivatives in the subtraction
change the first Friedmann equation (\ref{FE1}) from being algebraic in the 
Hubble parameter to containing second derivatives of it, and the particular
way (\ref{epsilonevo}) this change manifests is fatal for inflation. We were 
able to construct an analytic proof (\ref{smallepsprime}) --- supported by 
explicit numerical analysis in Figures~\ref{Fermiontry2} and \ref{Bosontry2} 
--- that the initial value of $\epsilon'$ can never be less than about 6. That
compares with its initial value of $5 \times 10^{-5}$ in the classical model,
and it means that inflation cannot last more than a single e-folding. So the
Ricci subtraction scheme is much worse than the initial time subtraction, but
neither method is satisfactory.

Before closing we should make a few comments. First, the problem with $\epsilon'(0)$
is almost completely independent of the classical model of inflation. So one
should not expect that a different model would lead to a different result. Second,
we need better control of the $\epsilon$ dependence of cosmological
Coleman-Weinberg potentials. The present study was carried out by assuming that
the constant Hubble parameter of de Sitter background becomes the instantaneous
Hubble parameter of an evolving geometry. In reality, the cosmological Coleman-Weinberg
potential should depend as well on $\epsilon(n)$ \cite{Kyriazis:2019xgj}. Accounting 
for this dependence will tighten the argument, and we expect it to extend the 
$\epsilon'(0)$ problem even to the initial time subtraction. Finally, it should be 
possible to extend these studies to the case in which derivatives of the inflaton 
are coupled to matter. Such a coupling would not induce a cosmological 
Coleman-Weinberg potential but would change the kinetic term. It would be very 
interesting to work out the consequences for evolution and the generation of 
perturbations. 

\vskip 1cm

\centerline{\bf Acknowledgements}

This work was partially supported by Taiwan MOST grants 
103-2112-M-006-001-MY3 and 107-2119-M-006-014; by the European 
Research Council under the European Union's Seventh Framework 
Programme (FP7/2007-2013)/ERC Grant No. 617656, ``Theories and 
Models of the Dark Sector: Dark Matter, Dark Energy and Gravity";
by NSF grants PHY-1806218 and PHY-1912484; and by the Institute 
for Fundamental Theory at the University of Florida.

\end{document}